\documentclass[pre,twocolumn]{revtex4}

\usepackage{graphicx}
\usepackage{verbatim}
\usepackage{comment}
\begin{document}

\def\K{{\bf{K}}}
\def\Q{{\bf{Q}}}
\def\Gbar{\bar{G}}
\def\epsbar{\bar{\epsilon}}
\def\Gc{{G_c}}
\def\tk{\tilde{\bf{k}}}
\def\k{{\bf{k}}}
\def\Na{{$N_\alpha$}}

\title{Cluster Solver for Dynamical Mean-Field Theory with Linear Scaling in
Inverse Temperature}
\author{E. Khatami,$^{1,2}$ C.~R.~Lee,$^3$ 
Z.~J.~Bai,$^4$ R. T. Scalettar,$^5$ and M. Jarrell$^2$} 
\affiliation{$^{1}$Department of Physics, University of Cincinnati, Cincinnati, Ohio, 45221, USA}
\affiliation{$^2$Department of Physics and Astronomy, 
Louisiana State University, Baton Rouge, Louisiana 70803, USA}
\affiliation{$^3$Computer Science Department, National Tsing Hua University, Taiwan}
\affiliation{$^4$Computer Science Department, 
University of California, Davis, California 95616, USA}
\affiliation{$^5$Physics Department, University of California, Davis, California 95616, USA}

\begin{abstract}
Dynamical mean-field theory and its cluster extensions provide a very
useful approach for examining phase transitions in model Hamiltonians, 
and, in combination with electronic structure theory, constitute 
powerful methods to treat strongly correlated materials.  The
key advantage to the technique is that, unlike competing real-space
methods, the sign problem is well controlled in the Hirsch-Fye (HF)
quantum Monte Carlo used as an exact cluster solver.  However, an
important computational bottleneck remains; the HF method scales as the
cube of the inverse temperature, $\beta$. This often makes simulations
at low temperatures extremely challenging. We present here a new method
based on determinant quantum Monte Carlo which scales linearly in
$\beta$, with a quadratic term that comes in to play for the number 
of time slices larger than hundred, and demonstrate that the sign 
problem is identical to HF.
\end{abstract}

\maketitle


\section*{INTRODUCTION}

Quantum Monte Carlo (QMC) methods provide an important methodology 
for solving for the properties of interacting Fermi systems. In 
auxiliary field techniques \cite{HHS,blankenbecler81,nightingale99,
silvestrelli93, zhang03,assaad99,tremblay92}, the partition function, 
$Z=\text{Tr}\ {\rm exp}[-\beta \hat H]$ is expressed as a path integral, for 
example, by discretizing the imaginary time $\beta$ into $L$ intervals 
of length $\Delta \tau$, and separating the one body (kinetic) and 
two-body (interaction) terms.  The latter are then decoupled through 
the introduction of a Hirsch-Hubbard-Stratonovich (HHS) field~\cite{HHS} 
which reduces the problem to a quadratic form.  The fermion degrees 
of freedom can be integrated out analytically, leaving an expression 
for the partition function which is a sum over the possible 
configurations of the auxiliary field.  For interacting lattice 
Hamiltonians, such as the Hubbard model, this field depends both upon 
the spatial site and on the imaginary time coordinate.  The sum over 
configurations is performed stochastically, for example, by suggesting 
local changes and accepting or rejecting with the Metropolis algorithm.  
The problem is challenging numerically because the summand is the 
determinant of a product of matrices, one for each fermion species. 
The determinant is costly to evaluate, and can also become negative 
at low temperatures, which constitutes the fermion sign problem~\cite{loh90}.

There are different ways to represent the matrices. In the determinant
quantum Monte Carlo (DQMC) approach \cite{blankenbecler81},  the
matrices have dimension equal to the number of spatial lattice sites
$N_c$. The matrices are dense, and involve the product of $L$ sparse
matrices. The algorithm scaling, $N_c^3 L$, arises from the need to update
$N_cL$ field variables at a cost of $N_c^2$ per update, where advantage is
taken of an identity for the inverse and determinants of $N_c$-dimensional
matrices which differ only by a rank-one change.  Simulations with this
method can now be done on many hundreds of spatial sites.  In situations
where particle-hole symmetry prevents a sign problem, for example, the
half-filled Hubbard Hamiltonian, one can reach arbitrarily low
temperatures. DQMC simulations have proven the existence of long-range
antiferromagnetic order in the two-dimensional half-filled Hubbard model
~\cite{hirsch89}, as well as accurately determined the nature of the
spectral function and thermodynamic properties at this 
density~\cite{paiva01,paris07}.

Alternatively, in the algorithm developed by Hirsch and Fye
(HF)~\cite{hirsch86} for embedded-cluster problems, a larger, sparse 
matrix of dimension $N_cL$ is considered.  The advantage of the HF-QMC approach 
is that the matrices are better conditioned (no product of $L$ matrices 
is involved) and also they remain positive to much lower temperatures; 
the sign problem is far less severe in the HF-QMC method.  However, because 
determinants of larger matrices are involved, the HF-QMC algorithm scales 
as $N_c^3 L^3$.  For this reason, HF-QMC has seen its most powerful applications
within dynamical mean-field theory (DMFT)~\cite{jarrell92,georges96} 
and its cluster extensions, the dynamical cluster approximation 
(DCA)~\cite{hettler:dca}, and the cellular dynamical mean-field theory
(CDMFT)~\cite{cdmft} for which $N_c$ is typically small.  In effect, DMFT 
trades the large lattice sizes $N_c$, and $N_c^3$ scaling of DQMC where 
spatial correlations can be explored, for the ability to reach much 
lower temperatures at general fillings at the cost of less 
real-space information, apart from that obtained from the mean field.  
DMFT also can directly access phase transitions which 
can only be inferred from finite-size scaling in DQMC.

In this paper, we describe a hybrid approach which combines some of  the
virtues of both DQMC and HF-QMC. The key algorithmic improvement is a
reduction in the $L^3$ HF-QMC scaling to linear in $L$.  The importance is
that this allows much larger $N_c$ to be considered.  At the same time, we
demonstrate analytically (and confirm numerically) that the fermion sign
problem in our hybrid algorithm is precisely the same as in HF-QMC, provided 
that the coupling to the host is fully taken into account.  Thus,
as in HF-QMC, we can reach low temperatures at quite general fillings.  Our
paper is organized as follows.  We first introduce the basic formalism,
including a proof that the sign problem is unchanged from HF-QMC.  We then
show results for various physical observables including the quasi-particle 
weight, local moment, and the Green's function. We demonstrate that the 
results of our algorithm converge to the same values as that of a
well-developed and tested HF-QMC code.  We conclude with a comparison of 
the scaling properties of our new approach.


\section*{FORMALISM}
\label{sec:formalism}

DMFT, DCA, and other cluster extensions such as the CDMFT all map the 
lattice problem onto an effective cluster embedded in a self-consistently 
determined effective medium. Here, we will add additional sites to the 
cluster to emulate the effective medium~\cite{m_caffarel_94,embedded}. 
The associated formalism will be sketched for the DMFT and DCA, but it is 
easily extendable to include CDMFT. 

The DCA is a cluster mean-field theory which maps the 
original $D-$dimensional lattice model onto a periodic cluster of size 
$N_c=L_c^D$ embedded in a self-consistent host.  This mapping is accomplished 
by replacing the  Green's function  and interaction used to calculate 
irreducible quantities such as the self-energy ($\Sigma$) by their coarse-grained 
analogs.  Spatial correlations up to a range $L_c$ are treated explicitly, 
while those at longer length scales are described at the mean-field level.
For details of the DCA formalism and algorithm, please see Ref.~\cite{maier05}.

The DCA loop converges when the cluster Green's function equals
the coarse-grained Green's function, $\Gc=\Gbar$,
\begin{eqnarray}
\Gbar(\K,i\omega_n) &=& \frac{N_c}{N_{t}} \sum_{\tk}
\frac{1}{i\omega_n-\epsilon_{\tk+\K} -\Sigma(\K,i\omega_n) } \\
&=&
\frac{1}{i\omega_n-\bar\epsilon_{\K} -\Sigma(\K,i\omega_n)-
\Gamma(\K,i\omega_n) }, 
\nonumber
\end{eqnarray}
where $\K$ labels a cluster wave number, $\omega_n$ is the Matsubara frequency, 
$\tk$ labels the lattice wave numbers 
in the Wigner-Seitz cell surrounding $\K$, and $N_t$ is the total number of 
lattice sites. $\bar\epsilon_{\K}=N_c/N_{t}\sum_{\tk}\epsilon_{\tk+\K}$ is the 
coarse-grained dispersion and $\Gamma$ is the single-particle hybridization 
between the DCA cluster and its effective medium.

Here, we consider the two-dimensional (2D) single-band 
Hubbard model~\footnote{Multiband 
models which involve only interband hybridization can be
easily treated in our method as in DQMC. In DQMC, density-density
interband/intersite interactions are known to produce a sign problem
which is significantly worse than onsite interactions. Spin-flip
(Hund's rule) type terms are even worse. This is also true in the
Hirsch-Fye approach. See K. Held, Ph.D. thesis, Universit\"{a}t Augsburg,
1999 (Shaker Verlag, Aachen, 1999).}. In order to employ DQMC 
as a cluster solver, we define an effective cluster Hamiltonian to preserve the 
coarse-grained Green's function through the addition of host band
degrees of freedom, which we label with $d^{\alpha}$.  
\begin{eqnarray}
H &=& \sum_{\K,\sigma} \epsbar (\K) c_{\K,\sigma}^\dagger c_{\K,\sigma}
+U\sum_i n_{i\uparrow} n_{i\downarrow}  \\
&+&\sum_{\K,\sigma,\alpha} \epsilon^\alpha(\K) 
d_{\K,\sigma}^{\alpha \dagger} d_{\K,\sigma}^{\alpha } 
+\sum_{\K,\sigma,\alpha} V_\K^\alpha c_{\K,\sigma}^\dagger
d_{\K,\sigma}^\alpha + \text{H.c.} \nonumber
\label{eq:Hamilt}
\end{eqnarray}
The host band label, $\alpha$, runs from $1$ to $N_\alpha$. $\epsilon^\alpha(\K)$ 
is the dispersion for the $d^{\alpha}$ band, $V_\K^\alpha$ is the coupling 
of the $d^{\alpha}$ band to the $c$ band, $U$ is the strength of the interaction 
and $n_{i\sigma}=c^{\dagger}_{i\sigma}c_{i\sigma}$ is the number of 
spin-$\sigma$ electrons on site $i$. Upon integration of the $d-$band
degrees of freedom, the correlated band Green's function becomes
\begin{equation}
G_{\text{eff}}(\K,i\omega_n) =\frac{1}{i\omega_n - \epsbar (\K) -
\Sigma(\K,i\omega_n)-\Gamma'(\K,i\omega_n)},\
\end{equation}
where
\begin{equation}
\Gamma'(\K,i\omega_n) =
\sum^{N_\alpha}_{\alpha=1}
\frac{\left|V_\K^\alpha\right|^2}
{i\omega_n-\epsilon^\alpha(\K)}.
\label{eq:gamma'}
\end{equation}
The parameters $V_\K^\alpha$ and $\epsilon^\alpha(\K)$ are adjusted 
to fit the DCA or DMFT hybridization function 
$\Gamma'(\K,i\omega_n)\approx \Gamma(\K,i\omega_n)$. 
For this, we use Marquardt's method~\cite{marquardt} to minimize the 
following merit function at each momentum point:
\begin{equation}
\chi^2(\K)=\sum_{n}\left|\Gamma(\K,i\omega_n)-\Gamma'(\K,i\omega_n)\right|^2.
\end{equation}
We define the scaled deviation as 
\begin{equation}
\eta(\K)=\frac{\chi(\K)}{\xi(\K)}
\label{eq:corr}
\end{equation}
where $\xi$ is the standard deviation of data. 

The discretization of the bath degrees of freedom has been 
considered in DMFT where exact diagonalization (ED) is used as the 
Hamiltonian-based impurity solver~\cite{m_caffarel_94}. Extensions of 
this method to dynamical cluster mean-field theories have also been largely 
implemented to study variety of models such as the extended Hubbard or 
multiband models~\cite{c_bolech_03,c_perroni_07,a_liebsch_09}. 
The advantage of this method is that since ED is essentially exact, there is no 
systematic error beyond the discretization of the bath. Moreover, more
complicated interactions than just the onsite Coulomb can be easily
included in the Hamiltonian.
However, the disadvantage of ED is that the Hilbert space grows exponentially 
with the total size of the system, $N_c(1+N_{\alpha})$. This greatly 
limits the size of the clusters that can be studied.
This is specially true since (as we discuss below) smaller clusters
generally require a higher number of non-interacting bands to fully account for the 
coupling to the bath, and for larger clusters, e.g.,
the 16-site cluster, even a very small $N_{\alpha} (=2)$ will make ED
inapplicable.


\section*{QMC ALGORITHMS AND THE SIGN PROBLEM}
\label{sec:sign}

The average sign in the DQMC method is equivalent to the average 
sign in the HF-QMC method in the limit of infinite number of bath bands, 
$N_\alpha\to\infty$. To prove this, we use the path-integral formalism and 
write the partition function as
\begin{equation}
\label{eq:Z-general}
Z=\int \mathcal{D}[\gamma]  \mathcal{D}[\gamma^*]e^{-S(\gamma,\gamma^*)}
\end{equation}
where $\mathcal{D}[..]$ denotes the functional integral, $S$ is the 
action, and $\gamma$ and $\gamma^*$ are Grassmann variable vectors. 
Equation (\ref{eq:Z-general}) can be approximated by 
\begin{equation}
\label{eq:Z-appr}
Z\approx \sum_{s_{i,l}=\pm1}\int \mathcal{D}[\gamma]  \mathcal{D}[\gamma^*]
e^{-S_0(\gamma,\gamma^*)} e^{-S_I(\gamma_c,\gamma_c^*)}
\end{equation}
where $S_{(0)I}$ 
is the (non)interacting part of the action and $\gamma_c$ and $\gamma^*_c$ 
represent the $c-$band components. In Eq. (\ref{eq:Z-appr}), we have used HHS
transformation to decouple the correlation in the interacting part 
of the action,
\begin{equation}
\label{eq:SI}
S_I(\gamma_c,\gamma_c^*)=-\sum_{i,l,\sigma} \lambda
\gamma^*_{c\ i,l,\sigma}\sigma s_{i,l}\gamma_{c\ i,l-1,\sigma}
\end{equation}
Here, $\cosh(\lambda)=e^{\Delta\tau U/2}$, $s_{i,l}$ is the auxiliary field 
and $l$ is the time index so that $\tau_{\ l}=l\Delta\tau=l\beta/L$. The 
non-interacting part of the action has the following form:
\begin{eqnarray}
\label{eq:S0}
S_0(\gamma,\gamma^*) &=& \Delta\tau \sum_{m,l,\sigma} 
\bigg[ \gamma^*_{m,l,\sigma} \left( \frac{\gamma_{m,l,\sigma}-
\gamma_{m,l-1,\sigma}}{\Delta\tau} \right) \nonumber \\ 
&+& H_0(\gamma_{m,l,\sigma},\gamma^*_{m,l,\sigma}) \bigg]
\end{eqnarray}
where $H_0$ is the non-interacting part of the Hamiltonian and $m$ denotes 
both the spacial coordinate and the band index (including the $c$ band). 
Equation (\ref{eq:Z-appr}) becomes exact in the limit of $\Delta\tau\rightarrow0$. 
By integrating out all the Grassmann variables in Eq. (\ref{eq:Z-appr}), one 
obtains the following expression:
\begin{equation}
\label{eq:Z-DQMC}
Z \propto \text{Tr}_{\{s_{i,l}\}} \det [G_{\uparrow}^{-1}] 
\det [G_{\downarrow}^{-1}]	
\end{equation}
where $G_{\sigma}$ is the Green's function of size $NL$
with $N=N_c+N_cN_{\alpha}$.

In the DQMC algorithm, $\Pi_\sigma \det[G_{\sigma}^{-1}]$ 
is used as the sampling weight to complete the sum over the auxiliary field. 
Note that the action is off-diagonal in 
time, except for the first term of the non-interacting action which is 
equal to one along the diagonal [see Eq. (\ref{eq:S0})]. Therefore, 
$G_{\sigma}^{-1}$ is an off-diagonal sparse matrix with identity matrices 
along the diagonal and its determinant can be evaluated from a smaller 
matrix of size $N$, using the following identity:
\begin{equation}
\label{eq:Bmatrix}
\det[G_{\sigma}^{-1}]=\det[I+B_{\sigma,L}B_{\sigma,L-1}\dots B_{\sigma,2}
B_{\sigma,1}]
\end{equation}
where $B_{\sigma,l}$ is the corresponding off-diagonal sub-matrix of 
$G_{\sigma}^{-1}$ at time slice $l$.  The DQMC Markov process proceeds 
by proposing changes in the HHS fields which are local in space and time, 
$s_{i,l}\to -s_{i,l}$.  Because of that, the ratio of the fermion 
determinants can be calculated directly from just the diagonal entry 
of the Green's function.  Similarly, the update of the Green's function 
following an accepted move does not require a full $\mathcal{O}(N^3)$ matrix 
inversion, but can be done in $\mathcal{O}(N^2)$ operations.  More details 
about this algorithm can be found in Ref.~\cite{blankenbecler81}.

Now suppose that instead of integrating out all the Grassmann variables 
in Eq. (\ref{eq:Z-appr}), we integrate out only the ones associated with the 
non-interacting electron bands. The partition function can then be written as 
\begin{equation}
\label{eq:Z-c}
Z \propto \sum_{s_{i,l}=\pm1}\int \mathcal{D}[\gamma_c]  \mathcal{D}[\gamma_c^*]
e^{-S_c(\gamma_c,\gamma_c^*)}
\end{equation}
where 
\begin{equation}
\label{eq:Sc}
S_c(\gamma_c,\gamma_c^*)=\sum_{i,l,j,l',\sigma}\gamma^*_{c i,l,\sigma}
\mathcal{G}^{-1}(i,l;j,l') \gamma_{c j,l',\sigma}+S_I(\gamma_c,\gamma_c^*).
\end{equation}
In the above equation, $\mathcal{G}$ is the non-interacting Green's 
function on the cluster ($\mathcal{G}^{-1}=G_{\text{eff}}^{-1}+\Sigma$) whose 
Fourier transform to momentum and frequency space can be written as 
\begin{equation}
\mathcal{G}(\K,i\omega_n) = (i\omega_n-\bar{\epsilon}_{\K}-
\Gamma'(\K,i\omega_n))^{-1}.
\end{equation}

In the limit of an infinite number of non-interacting host bands, 
$N_\alpha\to\infty$, the self-consistent DCA hybridization function may 
be exactly represented by the analytic form of Eq. (\ref{eq:gamma'}), 
$\Gamma'(\K,i\omega_n) =\Gamma(\K,i\omega_n)$. Therefore, $\mathcal{G}$ 
will be equal to the DCA cluster-excluded Green's function, $(\bar{G}^{-1}
+\Sigma)^{-1}$. By integrating out the rest of Grassmann variables in 
Eq. (\ref{eq:Z-c}), the partition function reads
\begin{equation}
\label{eq:Z-HF}
Z \propto \text{Tr}_{\{s_{i,l}\}} \det [G_{c\uparrow}^{-1}] 
\det [G_{c\downarrow}^{-1}]
\end{equation}
where $G_{c}$ is the DCA cluster Green's function of size $N_cL$.

In HF-QMC, to complete the sum over the auxiliary field, 
$\Pi_{\sigma}\det G_{c\sigma}^{-1}$ is used as the sampling weight. Unlike 
DQMC, where the inverse Green's function is sparse, here $G_c^{-1}$ is a dense 
matrix with a dimension that grows with the number of time slices. The HF-QMC 
Markov process proceeds by proposing local changes in the HHS fields, 
$s_{i,l}\to -s_{i,l}$. The cost to propose a change, i.e., to calculate the 
ratio of determinants [Eq. (\ref{eq:Z-HF})], is low and does not depend upon 
$L$ or $N_c$.  If a change is accepted, then the cluster Green's function 
matrix $G_c$ must be updated.  It is possible to write this step as a 
rank-one matrix update.  However, since the inverse Green's function matrix 
is dense, it is not possible to decompose it into $N_c\times N_c$ blocks 
similar to what was done above with DQMC.  

By comparing Eqs. (\ref{eq:Z-DQMC}) and (\ref{eq:Z-HF}), one 
can write the following equation for a particular field configuration:
\begin{equation}
C \det [G_{c\uparrow}^{-1}] \det [G_{c\downarrow}^{-1}]=
\det [G_{\uparrow}^{-1}] \det [G_{\downarrow}^{-1}].
\end{equation}
Since $C$ is independent of fields, the ratio of sampling weights will be 
the same and therefore, the measured quantities, including the average sign, 
will have the same statistics in DQMC and HF-QMC algorithms.


\section*{RESULTS}
\label{sec:results}

\begin{figure}[t]
\centerline {\includegraphics*[width=3.3in]{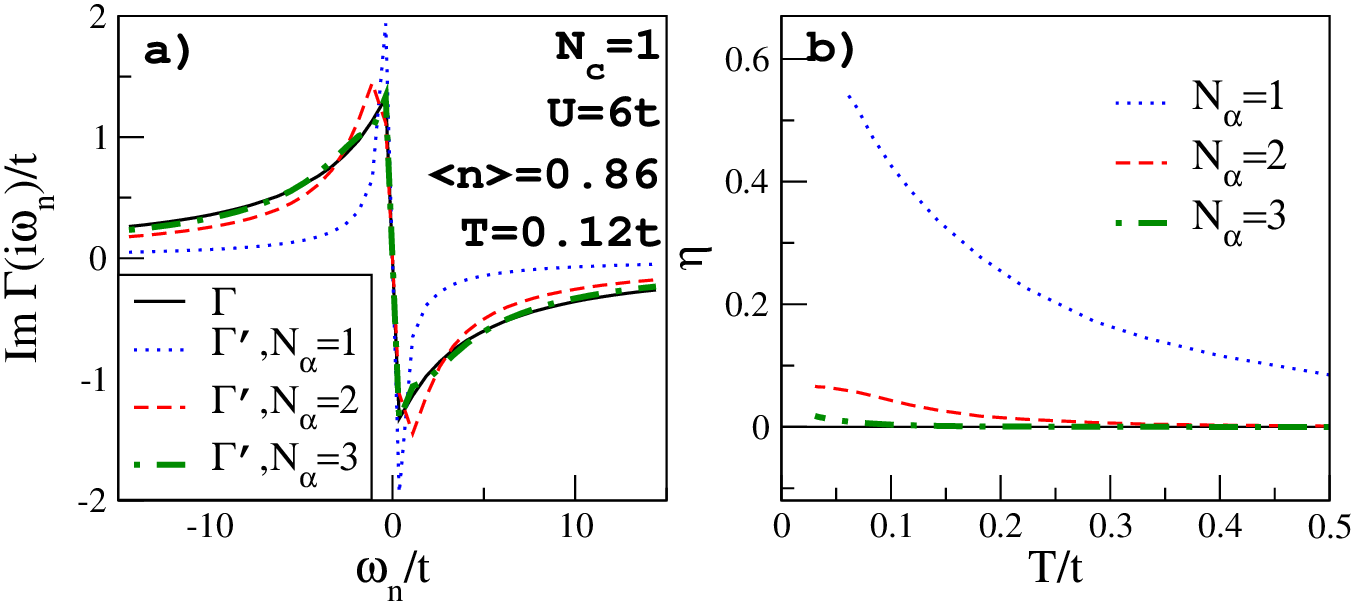}} 
\caption{(Color online) (a) The imaginary part of the DMFT hybridization 
function and fits to its analytic form 
of the effective cluster problem [Eq. (\ref{eq:gamma'})] for 
$N_{\alpha}=1, 2$, and $3$ versus Matsubara frequency. (b) The 
corresponding scaled deviations of the fits [Eq. (\ref{eq:corr})]
versus temperature.}
\label{fig:fitting}
\end{figure}

We apply this method to the 2D Hubbard model [Eq. (\ref{eq:Hamilt})] on
a square lattice with nearest-neighbor hopping, $t$, and show 
results for $\Delta\tau t=1/4$ and the interaction 
equal to three quarters of the bandwidth ($U=6t$) at filling, 
$\left<n\right>=0.86$, throughout this paper; calculations at 
different doping regions and for interaction strength equal to 
the bandwidth lead to the same trends for the quantities discussed in this work
\footnote{We have verified the reliability of this method in dealing with
systems with coexisting metallic and insulating solutions by reproducing the 
hysteresis curve in Fig.~2 of A. Macridin, M. Jarrell, and Th. Maier,
Phys. Rev. B {\bf 74}, 085104 (2006).}.
The quality of the fit of the effective cluster hybridization function 
[Eq. (\ref{eq:gamma'})] to the 
DCA or DMFT hybridization function, $\Gamma$, is improved by increasing 
the number of non-interacting bath bands. In Fig.~\ref{fig:fitting}(a), 
we show the imaginary part of $\Gamma(i\omega_n)$ and the 
corresponding data for $\Gamma'(i\omega_n)$ from the fitting algorithm 
using different values of $N_\alpha$ for a single impurity problem (DMFT). 
The improved quality of the fit at a 
low temperature ($T=0.12t$) can be seen as $N_\alpha$ increases from 
$1$ to $3$. We find that for a finite $N_\alpha$, the quality of the fit 
always decreases as the temperature is lowered. This can be seen in 
Fig.~\ref{fig:fitting}(b) where we show the scaled deviation 
of the fit [Eq. (\ref{eq:corr})] for different values of $N_\alpha$ as 
a function of temperature. The hybridization function is poorly fit for 
$N_\alpha=1$ even at high temperatures. However, the scaled deviation 
is strongly reduced when $N_\alpha$ increases. 

\begin{figure}[t]
\centerline {\includegraphics*[width=3.3in]{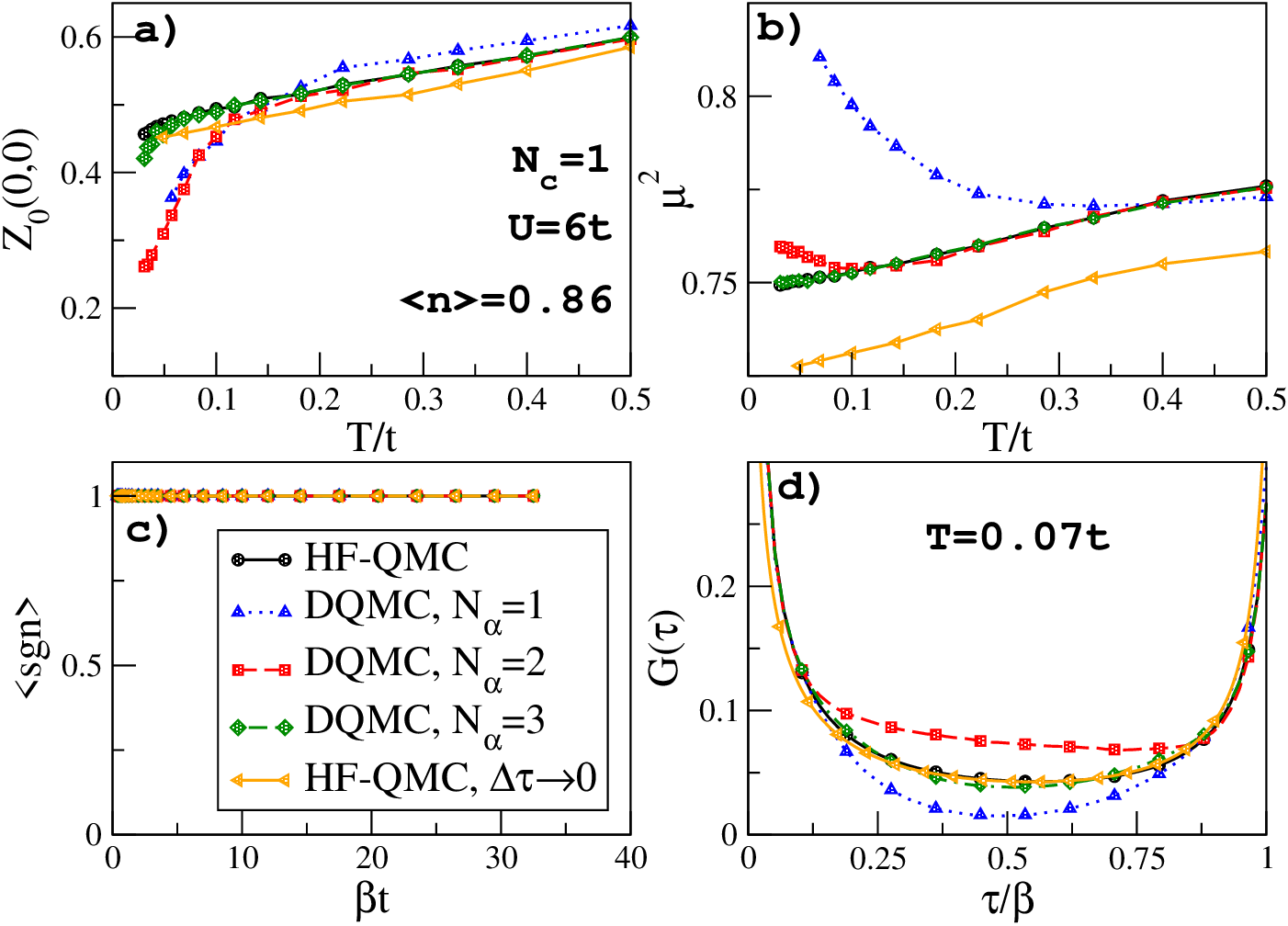}} 
\caption{(Color online) The convergence of DQMC to HF-QMC  
by increasing $N_{\alpha}$ for a single impurity problem (DMFT). We plot (a) the Matsubara frequency 
quasi-particle fraction versus temperature, (b) the unscreened moment versus temperature, 
(c) the average sign versus inverse temperature and (d) the Green's function 
at a low temperature versus imaginary time, calculated using HF-QMC and 
DQMC as impurity solvers. For comparison to exact results, a HF-QMC 
solution with very small $\Delta\tau$ is also presented. For DQMC, we show results for $N_\alpha=1,\ 2$,
and $3$. For a single-site problem, the average sign is exactly one in 
all cases. The statistical error bars are smaller than the symbols and are not shown.}
\label{fig:Nc1_n095}
\end{figure}

As the number of bath degrees of freedom increases, DQMC recovers the 
HF-QMC results for a single-site problem. We find that a maximum of 
four bath bands are sufficient for the agreement of the two methods 
at temperatures as low as $T=0.07t$. This convergence is shown in 
Fig.~\ref{fig:Nc1_n095} for $N_{\alpha}\leq 3$ where we plot the Matsubara 
frequency quasi-particle weight ($Z_0(\K)=[1-\text{Im}\Sigma(\K,i\pi T)/\pi T]^{-1}$), 
local moment ($\mu^2=\left<(n_{\sigma}-n_{-\sigma})^2\right>$) and the 
Green's function, calculated using HF-QMC and DQMC solvers. To have an
idea about the absolute errors, we have also included results from an exact 
solution, i.e., HF-QMC with a very small $\Delta\tau$ ($=1/16t$).
We point out that the average fermion sign, shown in Fig.~\ref{fig:Nc1_n095}(c), 
is equal to one, regardless of the bath in the single-site limit.

\begin{figure}[b]
\centerline {\includegraphics*[width=3.3in]{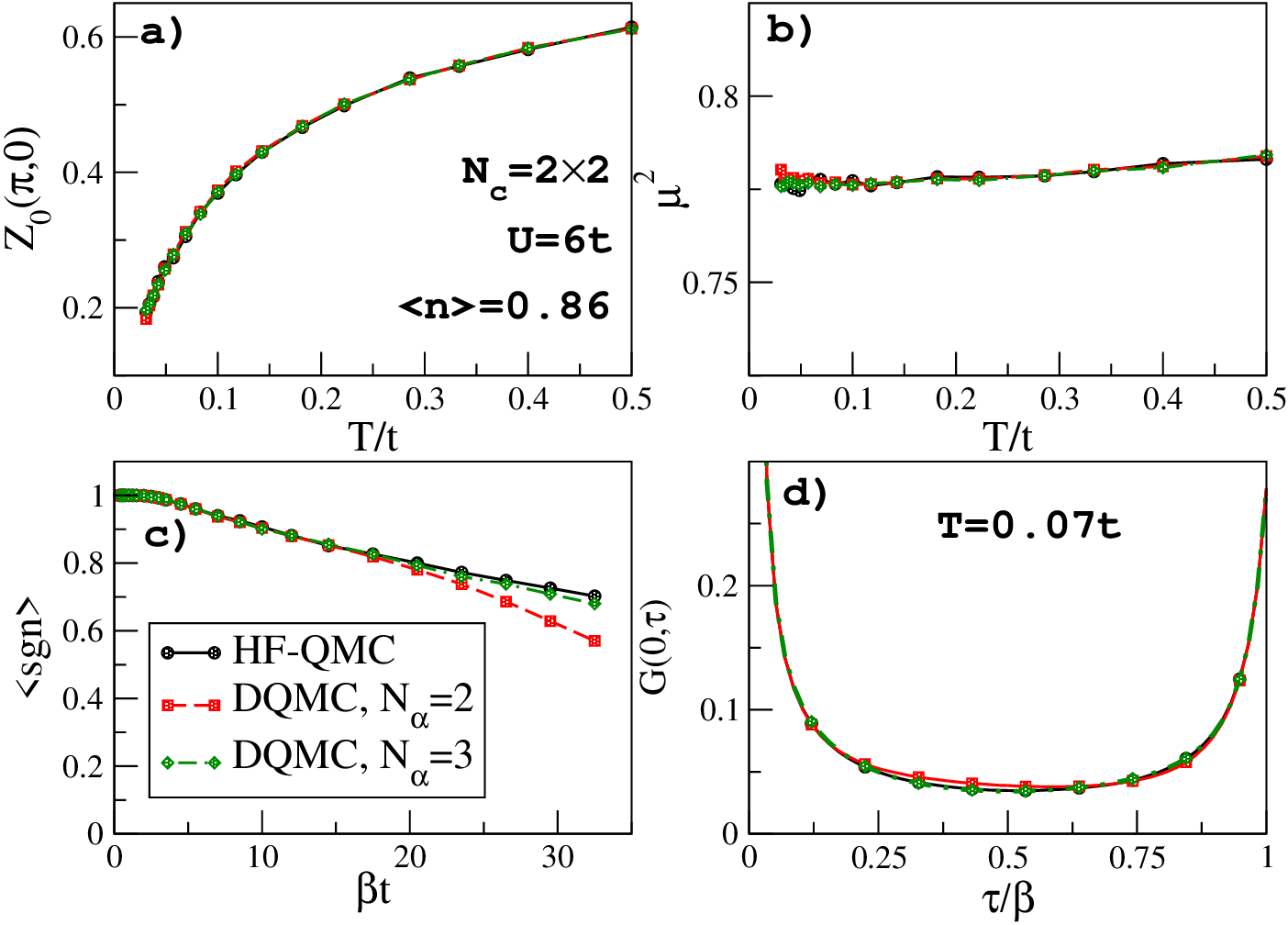}} 
\caption{(Color online) Same as Fig.~\ref{fig:Nc1_n095} for a $2\times2$ 
cluster in the DCA. In (a) and (d), we plot the quasi-particle fraction 
at $\K=(\pi,0)$ and the Green's function at the origin, respectively. 
Results for $N_{\alpha}=1$ cannot be obtained due to a bad sign, even at 
relatively high temperatures.}
\label{fig:Nc4_n095}
\end{figure}

The DQMC is a well-behaved cluster solver for the DCA as the number of bath 
bands needed to recover the HF-QMC results decreases with increasing cluster 
size. This can be understood from the suppression of the coupling between 
cluster and host degrees of freedom. In fact, it was shown previously that 
the hybridization function in the DCA is of order $\mathcal{O}(1/N_c^{2/D})$,
where $D$ is the dimensionality~\cite{th_maier_00}. To illustrate that, we 
plot in Fig.~\ref{fig:Nc4_n095} the same quantities of Fig.~\ref{fig:Nc1_n095} 
using the same model parameters but now calculated on a $2\times2$ cluster. For this 
cluster, the DQMC results show very good agreement with those of HF-QMC up 
to $\beta t=34$ when $N_\alpha=3$. As proven in the previous section, the 
average sign in DQMC converges to its HF-QMC value by increasing $N_\alpha$ [see 
Fig.~\ref{fig:Nc4_n095}(c)]. We find that the sign shows a strong sensitivity 
to the quality of the hybridization function fit. Thus, when $N_c>1$, results 
for $N_\alpha=1$ can not be obtained due to a bad sign problem, even at 
relatively high temperatures. In Figs.~\ref{fig:Nc4_n095}(a) and \ref{fig:Nc4_n095}(d), we show the 
quasi-particle fraction at $\K=(\pi,0)$ and the Green's function at the origin 
in real space, respectively.  

\begin{figure}[t]
\centerline {\includegraphics*[width=3.3in]{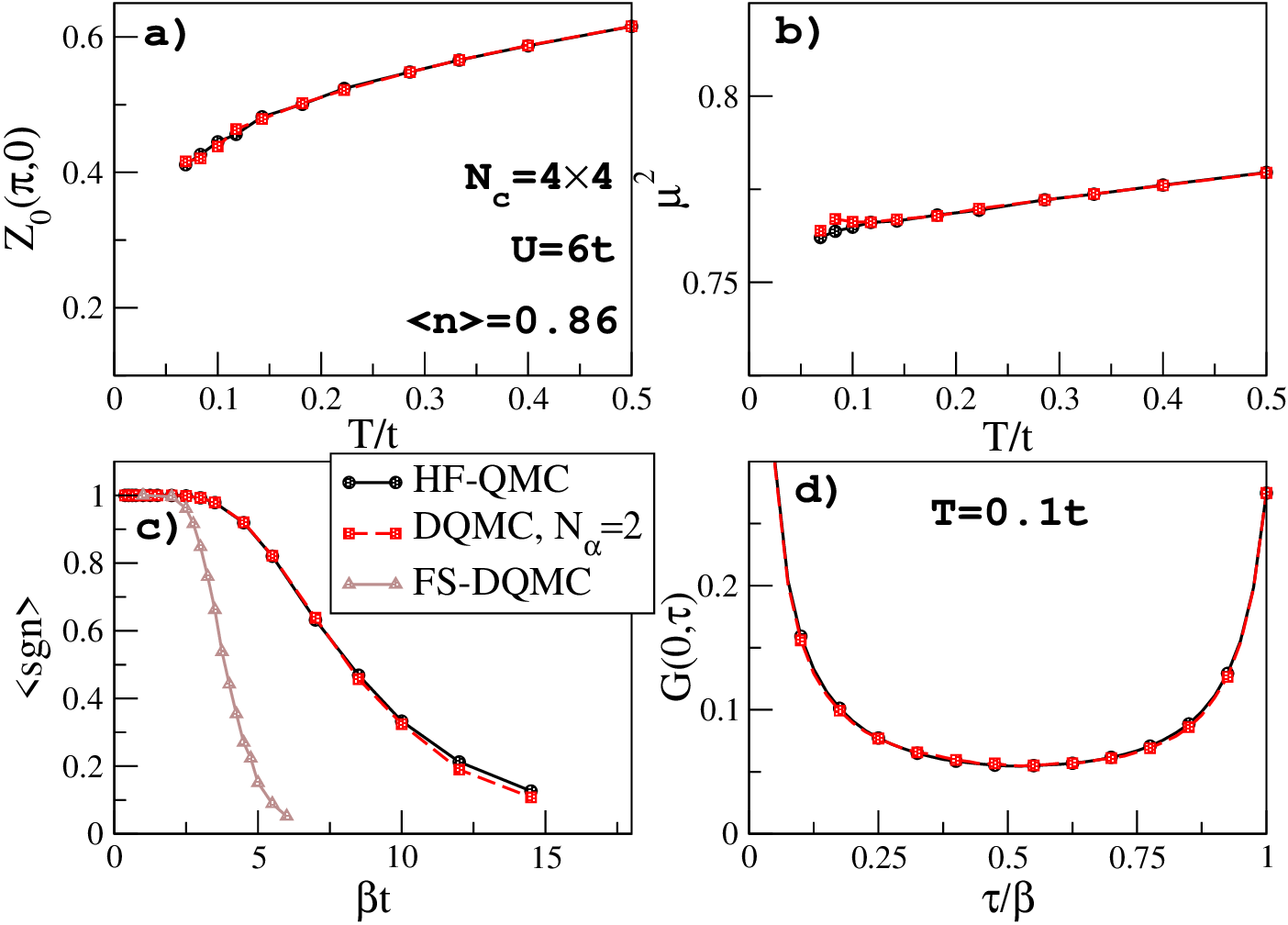}} 
\caption{(Color online) Same as Fig.~\ref{fig:Nc4_n095} for a $4\times4$ cluster. 
For this cluster, the convergence of DQMC to HF-QMC is achieved with $N_\alpha=2$.
In (c), we also show the average sign for a finite-size (FS) DQMC calculation
on this cluster using the same model parameters.}
\label{fig:Nc16_n086}
\end{figure}

The DQMC cluster solver is best suited for larger cluster simulations where 
$N_\alpha=2$ is sufficient to recover the HF-QMC results. As an example, 
we present results for a $4\times 4$ cluster in Fig.~\ref{fig:Nc16_n086}. 
We find excellent agreement between HF-QMC and DQMC calculations when 
$N_\alpha=2$. Here, the average sign falls more rapidly by decreasing 
temperature than that of the $2\times 2$ cluster [see Fig.~\ref{fig:Nc16_n086}(c)]. 
This limits the calculations for this cluster to $\beta t\leq15$ in 
the optimally doped region. However, as can be seen in Fig.~\ref{fig:Nc16_n086}(c), 
the average sign is significantly improved from a finite-size DQMC 
calculation. 

\begin{figure}[t]
\centerline {\includegraphics*[width=3.3in]{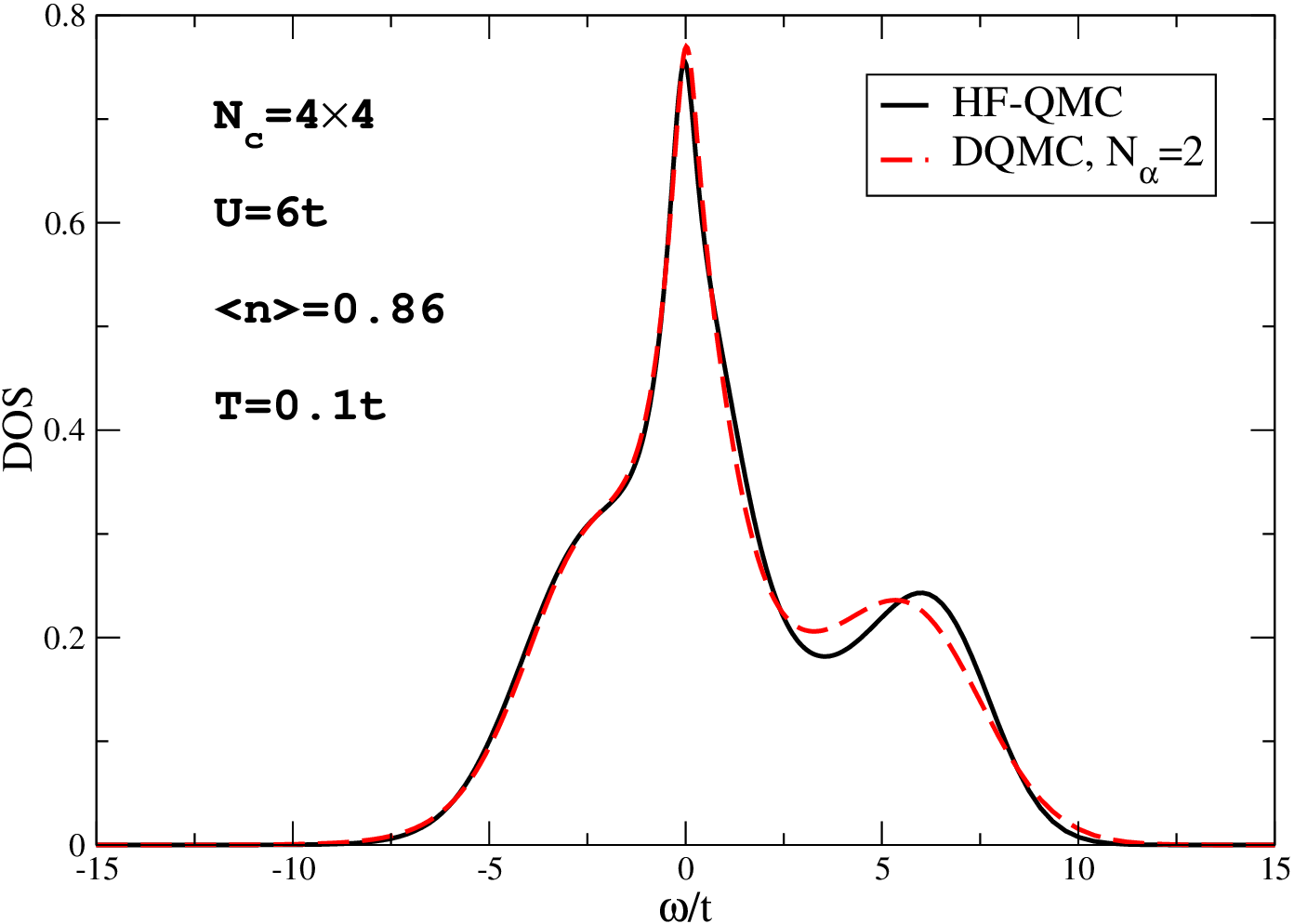}} 
\caption{(Color online) Density of states for the case study of 
Fig.~\ref{fig:Nc16_n086}. The solid (dashed) line shows the results for
HF-QMC (DQMC with $N_\alpha=2$).} 
\label{fig:DOC}
\end{figure}

As in HF-QMC, analytic continuation can be performed to 
calculate real-frequency quantities when DQMC is used as the cluster 
solver. As an example, we 
have considered the case of Fig.~\ref{fig:Nc16_n086} and calculated 
the single-particle density of states (DOS) using the maximum entropy 
method~\cite{mem}. The results indicate that discretizing the bath degrees
of freedom does not have a significant influence on the spectra. A comparison 
between HF-QMC and DQMC DOS has been presented in Fig.~\ref{fig:DOC} where 
we find that there is a very good agreement between the two density of 
states in the low energy region. However, there is a slight difference 
in the high-energy region which would presumably vanish by increasing 
$N_{\alpha}$.

\section*{SCALING}

\begin{figure}[t]
\centerline {\includegraphics*[width=3.3in]{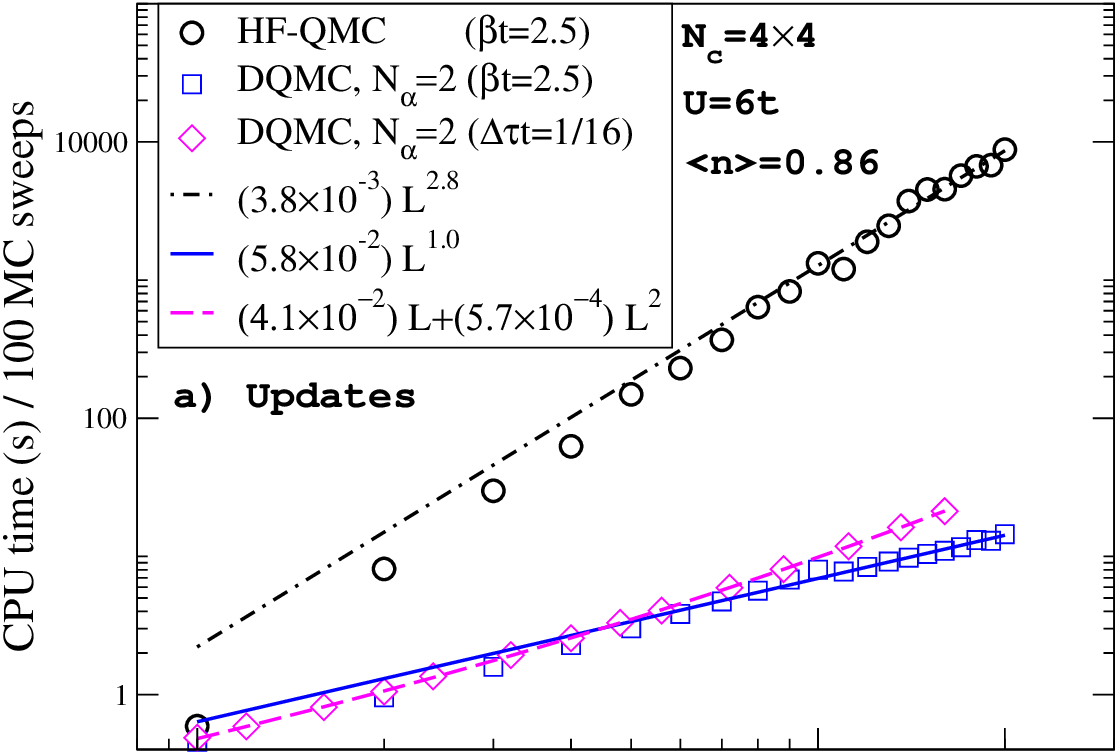}}
\centerline {\includegraphics*[width=3.3in]{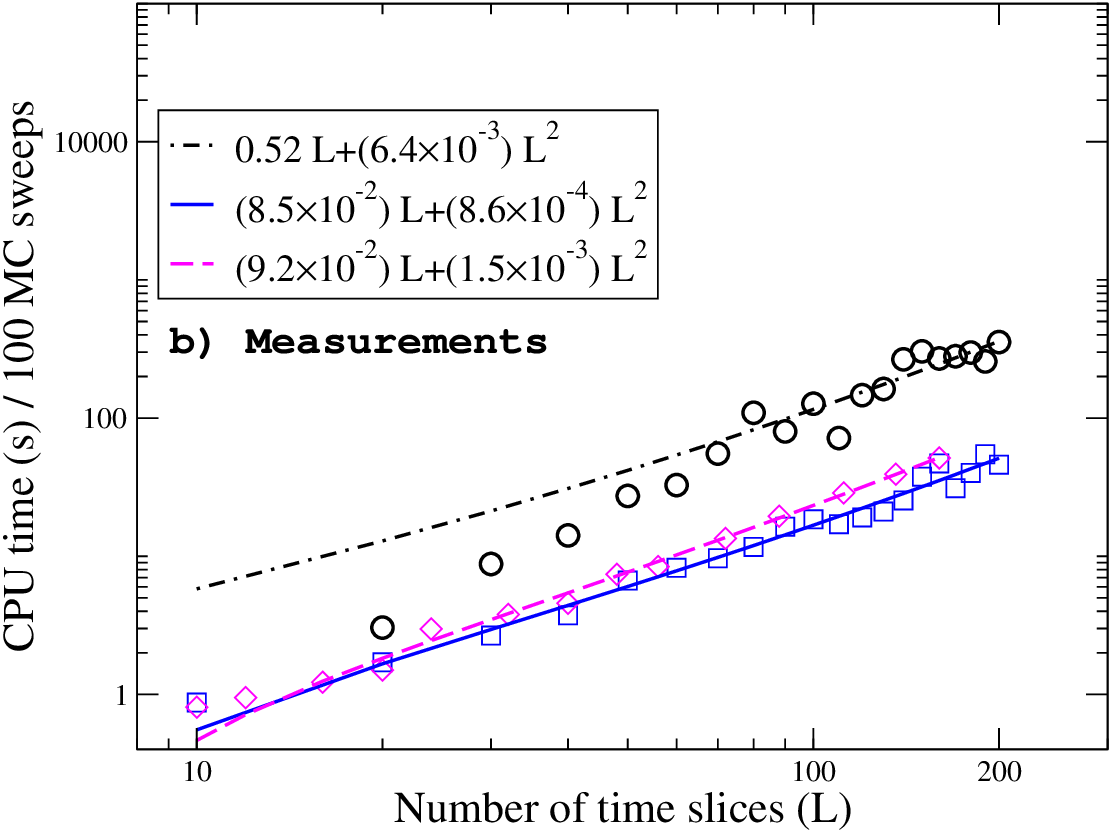}}
\caption{(Color online) The CPU time required for (a) updating 
and (b) measurement parts of 
the HF-QMC and DQMC ($N_\alpha=2$) algorithms versus the number of time 
slices on a $4\times4$ cluster. All other quantities are kept constant. 
The lines show power-law fits of the data. The diamond symbols 
show the CPU time in DQMC with a constant $\Delta \tau$ (decreasing 
temperature) where orthogonalization is performed to stabilize the matrix 
multiplications.}
\label{fig:update}
\end{figure}

As discussed in previous sections, the linear scaling of the DQMC 
algorithm with the number of time slices is the main advantage of 
this cluster solver over HF-QMC. The updating process in HF-QMC, 
which is the most expensive step in this algorithm, scales like 
$(N_cL)^3$. This is a results of $\mathcal{O}(N_cL)$ changes in the field 
variable during each sweep and $\mathcal{O}(N_c^2L^2)$ operations to update the 
Green's function for each change, using a rank-one updating mechanism. 
A similar argument applies to the scaling in the DQMC, except that it 
costs $\mathcal{O}((N_c+N_cN_{\alpha})^2)$ to update the inverse Green's function 
after each change in the field variable. Since the number of HHS fields
and therefore, the number of such updates is proportional to $L$, the 
overall scaling of updates in DQMC is linear in $L$. The scaling in the 
system size remains cubic as in other QMC methods and is a big advantage 
over ED which scales exponentially in the size. To show the linear 
behavior in $L$, we plot the CPU time for updates versus $L$ on the $4\times4$ 
cluster in Fig.~\ref{fig:update} (a). First, we compare this to that of HF-QMC 
for the same model parameters and by setting $\beta t=2.5$. At this 
fixed $\beta$, the product of matrices in DQMC is stable, which 
results in a perfectly linear scaling. We find that the updating 
step in DQMC is up to three orders of magnitude faster than in HF-QMC 
for a large number of time slices ($L\sim200$). 

In more realistic simulations, increasing $L$ is a consequence of 
increasing $\beta$ to access low temperatures for a fixed order of 
systematic error (constant $\Delta\tau$)~\cite{R_fye1,R_fye2}.
In this case, we do not expect to see any change in the scaling of HF-QMC. 
However, in DQMC, an orthogonalization step which scales as $L^2$, has to 
be performed to avoid the round-off errors. To show how the DQMC scaling 
changes, we also plot in Fig.~\ref{fig:update}(a), the CPU time for DQMC 
with $\Delta\tau t=1/16$. We see that the orthogonalization step 
introduces a quadratic term in $L$ with a coefficient which is two orders 
of magnitude smaller than the coefficient of the linear term [see diamond 
symbols in Fig.~\ref{fig:update}(a)]. This effect on the performance of the 
algorithm becomes slowly significant only when $L\geq100$. We point out that measuring 
the Green's function in DQMC involves matrix multiplications of the same 
type as in the updating process, and therefore results in the scaling 
of the CPU time that is very similar to the one for the updates. However,
as can be inferred from Fig.~\ref{fig:update}(b), measurements generally
take more time than updates and the quadratic term appears even
in the case of constant $\beta$. The time for measuring the Green's function in HF-QMC has 
more or less the same scaling as in DQMC, but is roughly an order of magnitude 
larger when $L \sim 200$.

\section*{DISCUSSION}

In this paper we have shown that the use of DQMC as a cluster solver
provides several order of magnitude speedup over the HF-QMC
algorithm, with a sign problem which is well behaved (identical to HF-QMC).
This improvement arises from a fundamental reduction in the scaling of
the algorithm, from cubic in the inverse temperature, $\beta$, to linear
in $\beta$ (with a small quadratic term arising from matrix
orthogonalization to reduce round-off errors).

However, the HF-QMC approach itself has already been supplanted in many 
applications by ``continuous time'' QMC (CTQMC) 
algorithms~\cite{rubtsov,assaad,rumbouts,karlis,p_werner_06,e_gull_07}.
We conclude this paper by addressing the relative strengths of the CTQMC 
technique and the new method presented here.  CTQMC eliminates the
systematic error inherent in HF-QMC and DQMC, including the method presented
here, by stochastically sampling the reducible Feynman graphs of the
partition function.  Although the matrix sizes are generally smaller
than in HF-QMC, the CTQMC algorithm also scales like the cube of the 
inverse temperature $\beta$~\cite{rubtsov}.  So, DQMC is generally much
faster than CTQMC when applied to finite sized systems~\cite{assaad} and
also for the embedded cluster problems presented here, especially at low 
temperatures. However, DQMC has the disadvantage of the introduction of 
systematic error.  These systematic errors in HF-QMC and DQMC may be 
eliminated by extrapolating the measured quantities in the time step squared,
$\Delta\tau^2\to 0$~\cite{blumer}. Since the values of $\Delta\tau$ that are 
used in this extrapolation are not overly small, the linear in $\beta$ 
nature of the present algorithm makes for far more efficient calculations, 
especially at lower temperatures.


\section*{CONCLUSIONS}
We have developed a DQMC cluster solver for the DMFT, DCA, or CDMFT which 
scales linearly in the inverse temperature but has the same minus sign problem
as HF-QMC.  Formally, this is accomplished by defining an effective Hamiltonian 
for the embedded-cluster problem which includes non-interacting bands for the host.  
The additional Hamiltonian parameters associated with the bath bands are adjusted
to fit the cluster-host hybridization function.  We prove that when this fit 
becomes accurate, this DQMC algorithm recovers the same average sign as HF-QMC.  
Using DCA simulations of the two-dimensional single-band Hubbard model, 
we demonstrate that as the number of bath bands increases, we recover 
the HF-QMC results, including the average sign.  The required number 
of bands is small, increases slightly with lowering temperature, and 
decreases with increasing cluster size.  

\section*{ACKNOWLEDGMENTS}  
We thank E.\ D'Azevedo, Simone Chiesa, and Karlis Mikelsons for stimulating 
conversations.  This work was funded by DOE SciDAC project, Grant No. DE-FC02-06ER25792
which supports the development of multiscale many-body formalism and
codes, including QUEST.  E.K. and M.J. were also funded by NSF Grant No. DMR-0706379.
This research was enabled by allocation of advanced  computing resources, 
supported by the National Science Foundation. The computations were performed 
on Lonestar at the Texas Advanced Computing  Center (TACC) under Account 
No.\ TG-DMR070031N, and on Glenn at the Ohio Supercomputer Center under 
Project No.\ PES0467.


\end{document}